# Ultra Low-Power SDM-based Circuit-Switching for Networks-On-Chip

Meysam Zaeemi, Mehdi Modarressi
School of Electrical and Computer Engineering, College of Engineering, University of Tehran, Tehran, Iran.
{zaeemi,modarressi}@ut.ac.ir

**Abstract.** In many modern AI chips and multicore systems-on-chip, embedded applications exhibit predictable inter-core traffic behavior that can be characterized at design time. For such applications, a variety of design-time traffic management and network optimization techniques can be employed to improve NoC power and performance. To exploit this predictability, we propose a novel low-power circuit-switched NoC design. It uses the Spatial Division Multiplexing (SDM) technique to establish circuits, implemented as subsets of NoC wires, for the communication flows of a target application. To further reduce the power profile of SDM, the design incorporates a new router architecture that combines hard-wired switches with conventional programmable crossbars. The architecture is complemented by an algorithm that maps application tasks onto a mesh NoC and assigns an SDM route with adequate bit-width to each circuit built for inter-task communication flows. Compared with a conventional packet-switched NoC, the proposed approach achieves approximately 38% lower NoC power consumption, 19% smaller area, and 12% lower packet latency.

## 1. INTRODUCTION

Network-on-Chip (NoC) architectures have become an integral component of modern many-core processors. Examples range from general-purpose many-core processors to complex systems-on-chip (SoCs) and specialized deep-learning accelerators, some of which integrate thousands of streamlined AI-based cores [1-3].

Beyond the traditional considerations of scalability and performance, power consumption has emerged as a first-order constraint in NoC design [4]. Consequently, designers must optimize NoCs not only for latency and delay but also for power efficiency.

Packet switching is the most widely used switching technique in NoCs because it provides high resource utilization and strong scalability under diverse traffic patterns [5]. It requires buffering incoming packets in every router and performing computations to determine routing paths, reserve buffer space in downstream routers, and arbitrate among packets competing for the same output port. These operations necessitate multistage pipelined routers, which in turn increase communication latency and energy consumption. Despite this overhead, the router's internal operations make it sufficiently intelligent to handle dynamically varying traffic patterns efficiently and scalable.

However, for most multicore embedded SoCs and large-scale AI chips [6-9], the target applications and their traffic characteristics are typically known at design time. Each application can be spatially partitioned into multiple tasks, with each task assigned to a dedicated processing element. As the cores execute fixed functions, the inter-core communication patterns tend to remain relatively static [10-11]. Designers can estimate the bandwidth requirements between any pair of nodes within an application through static code analysis or program profiling. In many such systems, inter-core communication occurs over a limited set of connections; that is, each task typically communicates with only a small subset of other cores [10].

Motivated by these observations, we propose using circuit-switching instead of packet-switching for such systems. The proposed design employs an SDM-based (spatial division multiplexing) approach to allocate NoC resources to circuits. In this scheme, the bit-width of links and other datapath components in the NoC is partitioned into several disjoint segments, with each segment allocated to a distinct circuit [12]. A dedicated path is established between the endpoint nodes of each communication flow for the lifetime of the application, with a bit-width proportional to the flow's bandwidth demand. Since packets travel along these dedicated paths, they require no buffering, routing, arbitration, or allocation operations, as they encounter no contention from other packets. Eliminating these time-consuming and power-intensive operations significantly reduces NoC latency and power consumption.

In this case, routers are reduced to very simple switches—namely, crossbar switches—that can establish arbitrary input-to-output connections between their ports through internal crosspoints. When using SDM, unlike a traditional crossbar that switches the full bit width of each input to a selected output port, the bit width is segmented, and each segment can be switched individually and independently of the others.

A chain of consecutive bit-segments in routers and links connected by internal router connections act as ad hoc dedicated virtual wires between the communicating nodes of the NoC, but reconfigurability is the key advantage of this virtual link over physical dedicated links.

n this paper, to further reduce the switch complexity and power overhead of internal crossbar crosspoints, some crosspoints are hardwired, statically and non-configurably connecting one incoming wire segment to an outgoing wire. In the circuit setup algorithm, if a free hardwired connection matches a circuit path, it is selected first for that circuit. The hardwired connections in this work are inspired by those previously proposed for FPGAs [12].

SDM-based resource partitioning in circuit-switched NoCs has been investigated in several prior studies [13-15]. It is generally considered a more power-efficient alternative to the widely used Time-Division Multiplexing (TDM) approach [16-19].

The main contributions of this paper are (1) proposing an efficient core-to-node mapping and flow routing (wire-to-flow allocation) algorithm and (2) adopting hard-wired switch connections to further decrease power consumption.

Finding a conflict-free set of paths for a given set of inter-node flows is an NP-hard problem that is efficiently solved by our design flow and algorithm. The proposed routing algorithm is developed based on the well-known *multi-commodity network flow* problem in graph theory.

The results show about 38% reduction in NoC power consumption, 19% smaller area, and 12% shorter packet latency.

## 2. SDM NoC Architecture

Figure 1 shows a router and a link in our design, where the entire NoC datapaths (buffers, links, and crossbars) are partitioned into several parts using the SDM scheme. Applying SDM allows us to have several parallel links with varying bit width in the same direction, each of which are part of a dedicated link for a communication flow between two communicating nodes.

Assuming the original link width is $N$ bits, the links are vertically partitioned into $N/m$ parallel sub-links (or units) each of $m$-bit width. Buffers and crossbars are also partitioned accordingly. In the crossbar switch, each $m$-bit unit of a crossbar input port can be connected to any $m$-bit unit at any crossbar output port. The number of wires per unit ($m$) is a design parameter ($1 \leq m \leq N$) that makes a trade-off between the router switching flexibility and switch area. If $m$=1, each individual wire of a crossbar input port can be connected to any wire at any crossbar output. On the other extreme, when $m$=$N$ (as in a conventional router crossbar), each wire of a crossbar input port can be only connected to the wire at the same position at some crossbar output.

The value of $m$ also determines the bandwidth allocation granularity. Smaller values of $m$ allow finer allocation granularities, which in turn results in more accurate bandwidth allocation, but at the price of larger crossbar switches. For example, consider a communication flow with 6 Mb/s bandwidth demand in a NoC where each individual wire provides 1Mb/s bandwidth. If $m$=1, six units are allocated to the flow which exactly match its bandwidth demand. However, if $m$=4, two units with a total bandwidth of 8 Mb/s should be allocated to cover the bandwidth demand of the flow that results in wasting some bandwidth. The latter, however, enjoys reduced crossbar cost and complexity.

A register at the input of each router buffers incoming data and sends it in the next cycle in a pipelined fashion to prevent NoC timing violation. In the conservative policy we use, data is buffered at each router. If the electrical characteristics of the links and router logic allows packets to pass more than one hop (router+link) in a single cycle (as it has been shown to be viable in some previous work [20]) buffering can be done once every multiple hops to further reduce latency.

Once a circuit with proper bit-width is allocated to a communication flow, the network interfaces at the endpoint nodes serialize/deserialize data into data units (or flits) of the same width as the allocated circuit width in an end-to-end basis. This process can be easily done by shift registers.

In the proposed NoC, packets move on dedicated reconfigurable paths and traverse one NoC hop per cycle. Simple router design and conflict free paths results in very low-latency and low-power communication, compared to a conventional packet-switched router where forwarding a packet is done after a sequence of time consuming and power-hungry buffering, routing, arbitration and flow control operations.

The power-efficiency of the proposed NoC can be further improved by replacing a portion of programmable crossbar cross-points by hard-wired fixed connections as shown in Figure 1. A fixed hard-wired cross-point removes the parasitic capacitance and resistance of a configurable switch, hence trades off switch reconfigurability with power-efficiency and performance. Utilizing a mixture of hard-wired and conventional programmable switches, we first consider hard-wired resources to build a path for communication flows and only use regular switch cross-points when hard-wired resources are not available. Wang et al. show that the use of hard-wired connections inside FPGA switch boxes considerably reduces circuit delay, area and power dissipation [12]. To the best of our knowledge, our proposal is the first to use the hard-wired crossbar connections in NoC routers.

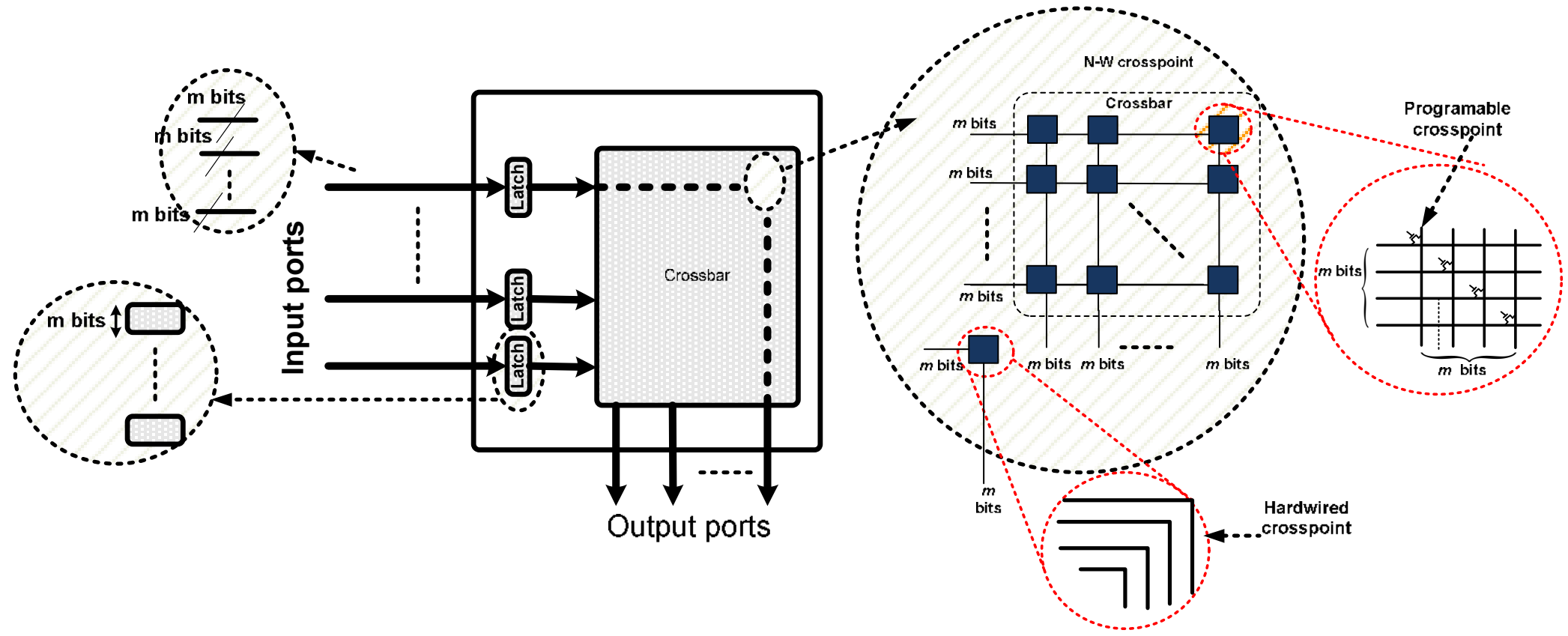

**Figure 1. The architecture of a SDM router with the details of a crossbar cross-point.**

In a synchronous NoC, hard-wired connections have no impact on communication latency, because the NoC clock is bound by the slower reconfigurable connections. However, they greatly reduce NoC power consumption, as it provides low-power switch-free paths for some circuits.

The crossbar switch of the proposed router is larger than of a conventional packet-switched router, because it has more cross-points. However, removing buffers and routing and arbitration logic not only compensate for the larger crossbar, but also reduce the total NoC area. Synthesis results of the proposed router (128-bit links, m=8) by a commercial tool show that it consumes 19% less logic than a conventional packet-switched NoC router (128-bit links, 8-entry buffers) in 65nm technology. Adding 25% hard-wired cross-points further increases area saving to 23%.

# 3. SDM NoC Design Flow

Each input application in our flow is described by a communication task-graph (CTG). CTG is a directed graph, $G(V,E)$, where each vertex $v€V$ represents a task (or respective core) and each directed edge $e_{i,j}€E$ represents a communication flow between tasks *i* and *j*. Each edge is tagged by the bandwidth demand of the flow it represents.

Given an input CTG, a NoC with mesh topology, N-bit links arranged as *N/m m*-bit units, *L* hard-wired connections (out of N) in each port, and a mapping of tasks onto NoC nodes, the algorithm finds a route for each flow in such a way that (1) routes are laid out along one of the shortest paths between the source and destination nodes of the flow, (2) contains the proper number of wire-units to satisfy the bandwidth requirement of the flow, (3) the cumulative width of routes assigned to a link do not exceed the link bandwidth , and (4) routes have no conflict (i.e. no common wire-units) with other circuits.

We first determine the number of units required by each flow which is calculated based on the bandwidth of each unit (bandwidth of a single wire multiplied by the number of wires per unit), and the bandwidth requirement of the flow. To improve the efficiency of the algorithm, we allow multi-path routing that allocates wire-units along different paths to a flow, when there are not sufficient unallocated wire-units along a single path.

For example, a flow that requires a 5-bit circuit can be provided by a 3-bit and a 2-bit circuit along different paths, if there are not sufficient resources to construct a 5-bit path. Travelling on paths with the same length and also experiencing no blocking, those parts of a packet that are sent along different paths reach the destination node at the same time.

As there are some strong similarities between this problem and the *multi-commodity network flow* problem in graph theory, we map our problem to a network flow problem and take advantage of its efficient solvers to find proper SDM route for input application flows.

The multi-commodity network flow problem is defined as follows. A flow network is a directed graph of nodes connected by arcs where each arc has a capacity, as well as a cost associated with it. There are also some traffic flowing in the network among some source-destination pairs that may go through multiple hops (and consume arc capacity at each hop) along its route in order to get from source to destination. The traffic capacity demands can be split onto different routes, i.e., the complete traffic is not necessarily routed through a single path. The goal of the problem is to find the set of routes with minimum cost through the network for each of those demands, in such a way that the total flow along each arc must be less than the arc's capacity.

We can easily map our problem to the above network flow problem. To this end, each NoC node and link is considered as a network node and arc, respectively. The capacity of each arc is set to the available bandwidth of its corresponding link in terms of the number of wire units. Communication flows (CTG edges) and their bandwidth demand (in terms of number of units) is modeled by traffics and their capacity demand in the network flow problem. We set the cost of all arcs (equivalent to NoC links) to the same value. However, to encourage flows to use hard-wired cross-points, we insert an arc with smaller cost than a regular link for each part of the links that are connected to hard-wired connections. Moreover, we limit the algorithm to a search inside the rectangular region between the source and destination nodes of a flow in order to guarantee that the found paths are minimal (in terms of hop count).

Once the problem is mapped to a multi-commodity network flow problem, we use a well-known solver, AMPL-CPLEX package [21], to solve the problem and find a route for every CTG flow. The tool can find appropriate solutions for all benchmarks listed in Section 4 in an acceptable time.

An important issue in the design flow is the CTG task to NoC node mapping that determines the place of each CTG task in the NoC. For this step that should be done before finding route for flows, we use a good heuristic algorithm that aims to minimize the total on-chip traffic by minimizing:

$$Min\{\sum_{\forall e_{i,j}} t(e_{i,j}) \times dist(M(v_i), M(v_j))\}$$

Where $dist(a,b)$ is the Manhattan distance between nodes $a$ and $b$ in the network and $M(v_i)$ is the network node to which CTG node $v_i$ is mapped. $t(e_{i,j})$ is the bandwidth demand of CTG edge $e_{i,j}$. NMAP guarantees that by the found mapping, all flows can find enough bandwidth to send data to the other endpoint node of their communication flows and the bandwidth limit of no link will be violated. Although NMAP is a mapping and routing mechanism designed for packet-switched networks, we just use its mapping scheme and then apply our SDM scheme to find a connection for every CTG flow.

We omit the details of the algorithm due to the limited space and refer interested readers to our previous work [10] for more details.

# 4. Experimental Results

In this section, we compare the power/performance of the proposed NoC architecture and design flow with a packet-switched and a conventional SDM NoC. The selected topology for all NoCs is a 2-dimentional mesh and all NoCs use the same size and core mapping using NMAP. The NoC bit width and packet size are set to 128 and 1024 bits, respectively. Consequently, each packet is divided into 8 flits[1].

The packet-switched NoC uses wormhole routers with 8-entry buffers per each input port. It adopts the well-known dimension-order (XY) routing algorithm with the look-ahead routing technique [15]. The conventional SDM NoC we selected for the comparison purpose, implements SDM resource partitioning, but uses a heuristic routing algorithm and does not utilize hardwire connections [22].

We developed a simulator for the SDM-based NoCs, but the packet-switched NoC is simulated using BookSim, a widely-used cycle-accurate simulator for interconnection networks [23]. The power consumption of all considered NoCs are calculated using an accurate architecture-level power model for NoCs in 45 nm.

The considered NoCs are evaluated under a set existing SoC designs that are widely-used as benchmark in the literature, including Multi-Window Display (MWD) with 13 tasks and 15 communication flows [24] mapped into a 4×4 mesh, Video Object Plane Decoder (VOPD) with 16 tasks and 21 flows [24] mapped into a 4×4 mesh, Multi Media System (MMS) with 27 tasks and 36 flows [24] mapped into a 5×6 mesh, GSM decoder (GSM-dec) with 48 tasks and 73 flows [25] mapped into a 7×7 mesh, GSM encoder (GSM-enc) with 36 tasks and 56 flows [25] mapped into a 6×6 mesh, Robot (from the STG suite [26]) with 81 tasks and 118 flows mapped into a 9×9 mesh. We also used a combination of two benchmarks from the E3D suite [27], namely Telecom with 24 tasks and 25 flows mapped into a 6×4 mesh, and auto-industry with 22 tasks and 25 flows mapped into a 6×4 mesh.

Figure 2 shows the average packet latency and power consumption of the proposed and a packet-switched NoC. Out of 128 links of our NoC, 48 links are connected to hard-wired and 80 links are connected to reconfigurable cross-points. The links in our design is divided into 32 4-bit units. As the figure shows, our proposed architecture and design flow effectively reduce the NoC power consumption and average packet latency by up to 17% (12% on average) and 47% (38% on average), respectively.

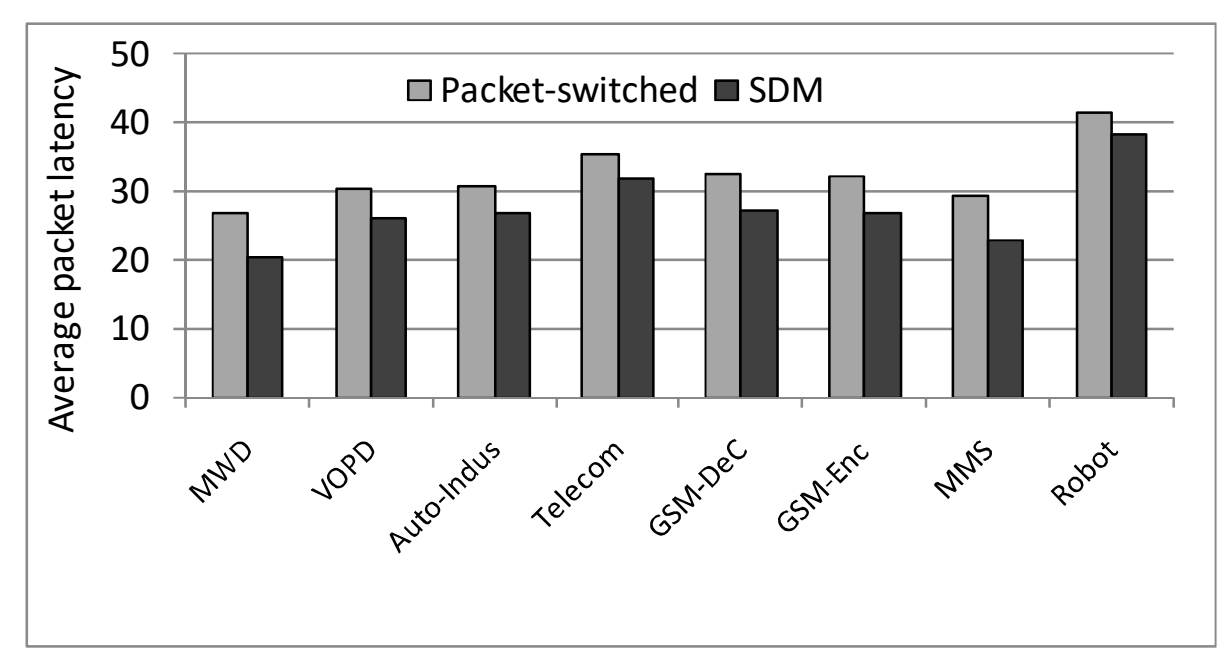

**(a)**

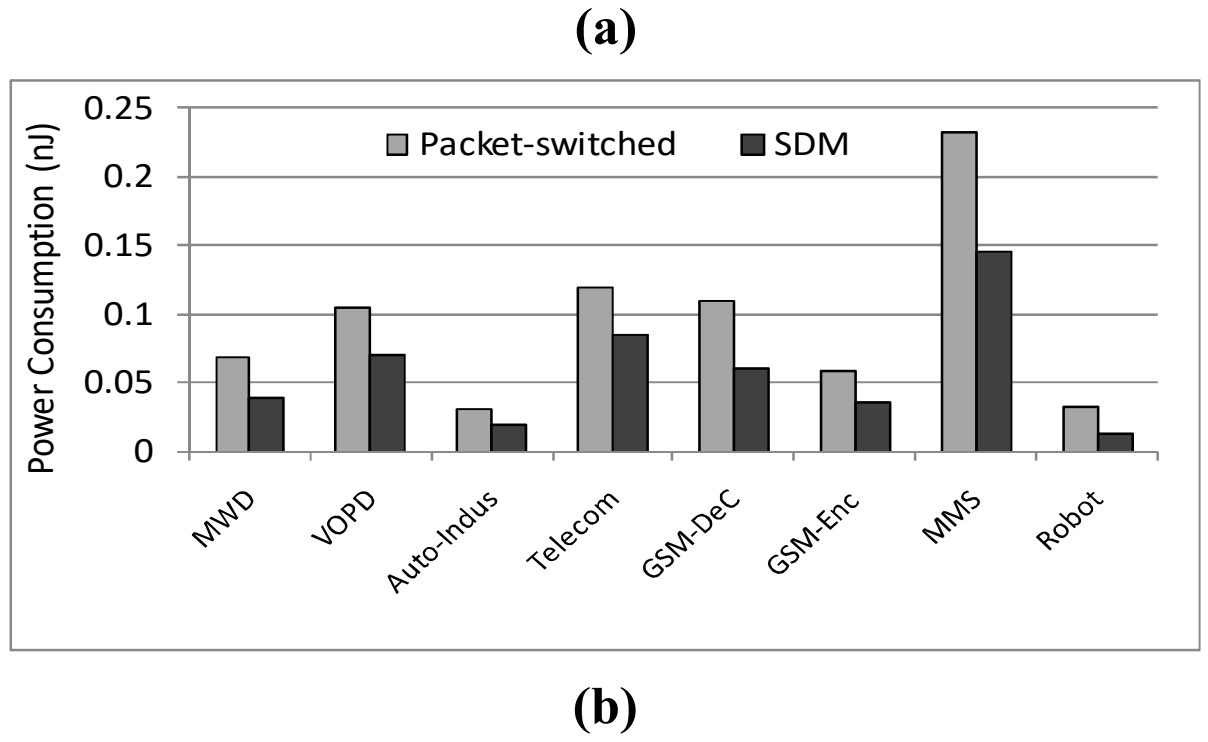

**(b)**

**Figure 2. The average packet latency (a) and power consumption (b) of the proposed and a packet-switched NoC under the considered benchmarks**

[1] Flit (flow control digit) is a unit of data into which a packet is broken in packet-switched NoCs. Generally, the flit size is equal to link bit-width, allowing the entire flit to pass a link (and other datapaths) at the same time.

Our design also uses smaller routers as discussed in Section 2. Please note that we set the frequency of each NoC proportional to the bandwidth demand of each benchmark, in order to enable the NoC to work in normal conditions (below saturation point) under the benchmark. For fair comparison, the frequency of both SDM and packet-switched NoCs is set to the same value when running the same benchmark.

Figure 3 shows the effect of hard-wired connections on NoC power consumption. As the figure shows, by connecting 48 bits of each link to hard-wired cross-points, more than 14% power saving can be achieved. Nonetheless, further simulations showed that using more hard-wired connection reduces the flexibility of the crossbar and may lead to unsuccessful routing for some flows. In this case, flows may not find enough resources while there are free hard-wired connections to other directions.

To evaluate the effectiveness of the proposed routing algorithm (path finding and bit-width allocation), we compare our results with the results obtained by the heuristic algorithm proposed in [7]. It uses a greedy algorithm that first sorts flows in the decreasing order of a metric composed of the flow bandwidth demand and routing flexibility (number of available paths between the source and destination nodes of the flow). Then, each flow is considered in the order and proper number of wires is reserved for it by examining all possible shortest paths between the flow endpoint nodes.

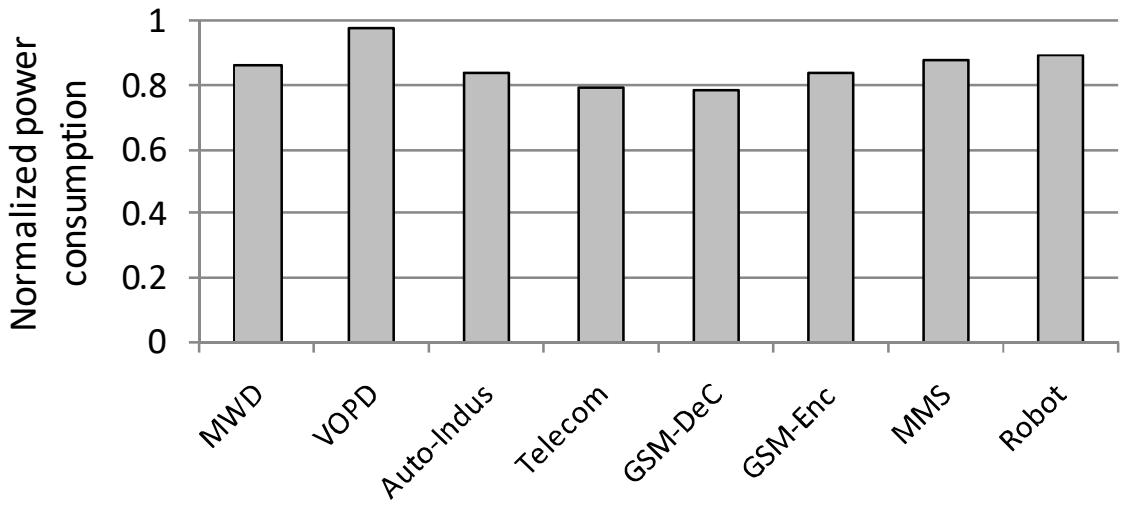

**Figure 3. Power reduction of SDM when 48 bits of each port are connected to hard-wired connections. The numbers are normalized to the results of baseline SDM**

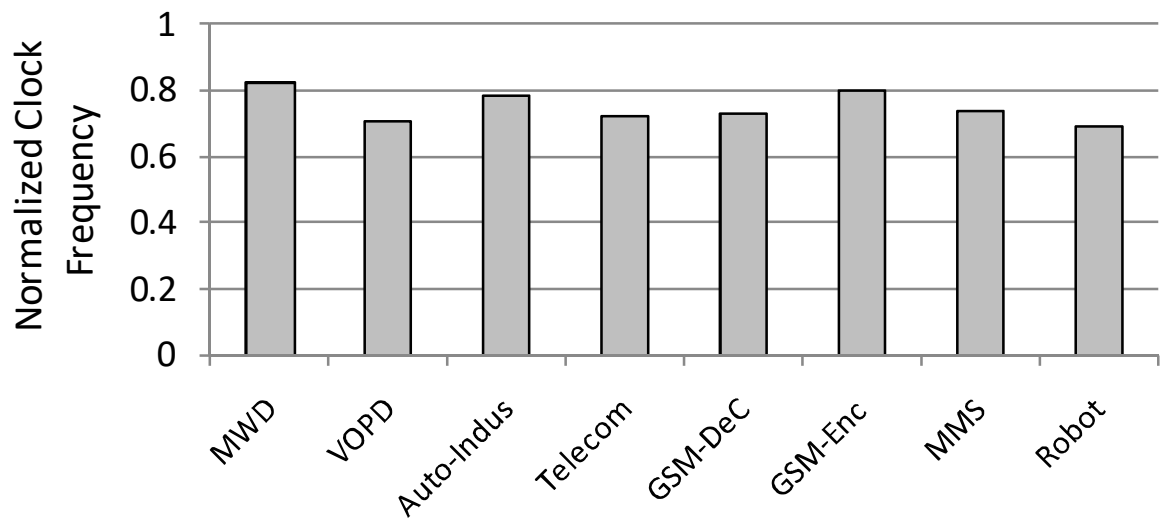

**Figure 4. Comparing the proposed algorithm with the algorithm in [7] in terms of the clock frequency at which the algorithm can find a path for all flows. Each bar shows the results of our algorithm normalized to the results of [7]**

In this experiment, if an algorithm cannot route all flows in a certain frequency and link bit-width, the frequency of the NoC is increased in order to increase the bandwidth of each wire. The flows, then, need less number of wires and it is more likely to find a valid route for all flows.

As Figure 4 shows, our algorithm can efficiently exploit existing NoC resources, and hence can find a routing at lower frequencies (27% on average) than of the other considered method.

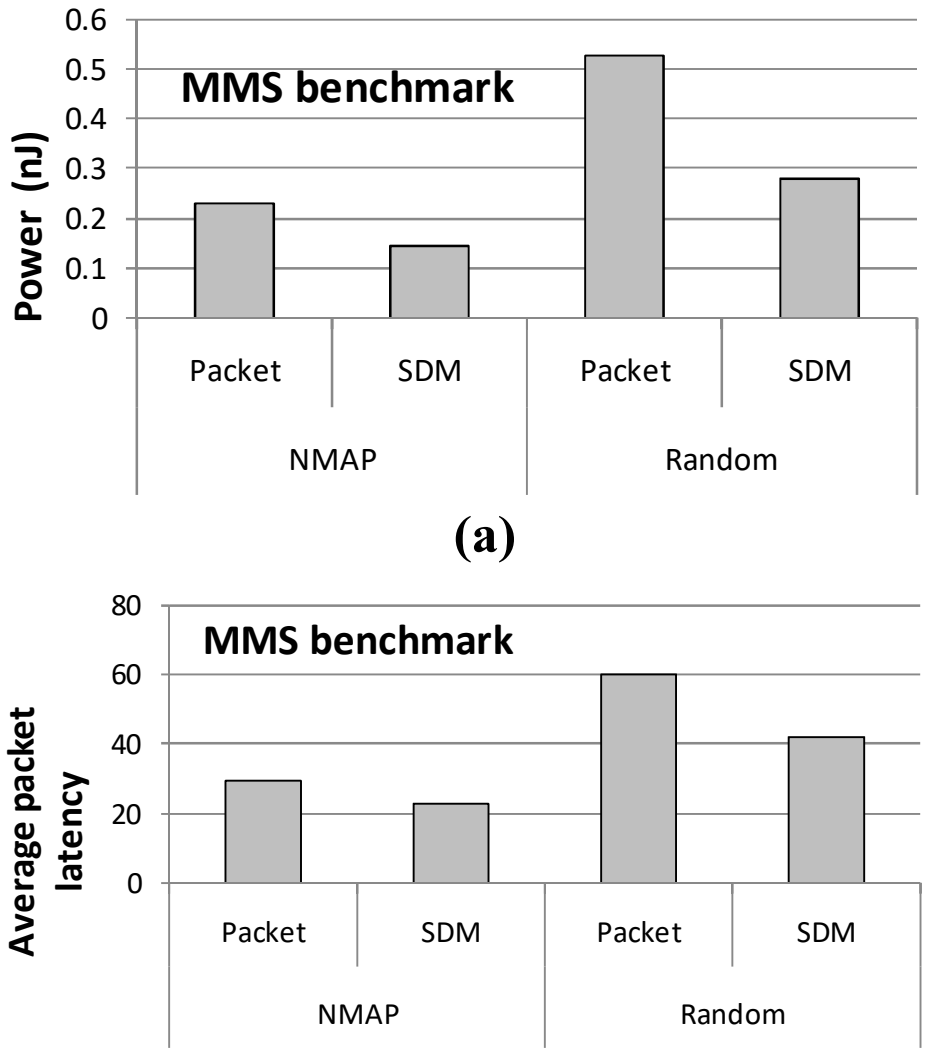

**Figure 5. The effect of mapping on the obtained (a) latency and (b) power improvements**

Finally, Figure 5 show the effect of mapping on the achieved power/performance gain on MMS benchmark. In these experiments, we compare the power/performance of a packet-switched and SDM-based NoCs with the mapping mechanism mentioned in Section 3 and a random mapping. Having random mapping for tasks is a realistic scenario and occurs when a new application is introduced for the chip after the physical mapping of cores is done. In this case, the placement of the cores is not optimized for the application and hence, the power and latency of communication increases. Unoptimized mapping leaves more opportunity for our method to optimize power/performance, as can be seen in Figure 5.

# 5. Conclusion

In this paper, we proposed an SDM-based NoC architecture as a more power-efficient alternative for packet-switched NoCs. The design targets NoC-based multicore SoCs used in embedded systems where the traffic pattern of applications can be characterized a priori. The simple routers of the proposed NoC utilize a set of hard-wired connections together with the conventional configurable crossbar in order to further reduce on-chip communication power consumption. We then developed a mapping and routing algorithm to distribute the NoC bit width (at links, buffers, and crossbars) among the communication flows of a running application. The size of the partition devoted to each communication flow is proportional to its bandwidth demand. Constructed based on the network flow problem in graph theory, the routing algorithm can efficiently find a path with appropriate width for every flow in all considered benchmarks. Compared to a conventional packet-switched NoC, our results show considerable reduction in NoC power consumption, latency, and area. The results show that this simple low-power NoC is an appropriate alternative of intelligent packet-switched NoCs in most embedded systems, where the traffic pattern of applications can be pre-characterized.

# References

[1] Z. Jin *et al.*, "Uncovering Real GPU NoC Characteristics: Implications on Interconnect Architecture," *2024 57th IEEE/ACM International Symposium on Microarchitecture (MICRO)*, 2024.

[2] Y. Yuan, et al., "Intel accelerators ecosystem: an SoC-oriented perspective: Industry product," *ACM/IEEE 51st Annual International Symposium on Computer Architecture (ISCA)*, 2024.

[3] S. Xiao, et al., "Neuronlink: An efficient chip-to-chip interconnect for large-scale neural network accelerators," *IEEE Transactions on Very Large Scale Integration (VLSI) Systems* 28.9, 2020.

[4] A. A. Kedilaya *et al.*, "Beyond backside power: backside signal routing as technology booster for standard cell scaling," *IEEE Journal on Exploratory Solid-State Computational Devices and Circuits*, 2025.

[5] W. J. Dally, and B. Towles, Principles and practices of interconnection networks, Morgan-Kaufmann Publishers, 2004.

[6] M. Musavi, E. Irabor, A. Das, E. Alarcón and S. Abadal, "Communication Characterization of AI Workloads for Large-scale Multi-chiplet Accelerators," *in Proc. ISCAS*, 2025.

[7] A. Firuzan *et al.*, "Reconfigurable Network-on-Chip for 3D Neural Network Accelerators," in *12th IEEE/ACM International Symposium on Networks-on-Chip (NOCS)*, Torino, Italy, 2018.

[8] R. Hojabr, M. Modarressi, M. Daneshtalab, A. Yasoubi, and A. Khonsari, "Customizing Clos Network-on-Chip for Neural Networks," *IEEE Transactions on Computers*, vol. 66, no. 11, pp. 1865–1877, Nov. 2017.

[9] N. Akbari and M. Modarressi, "A High-Performance Network-on-Chip Topology for Neuromorphic Architectures," in *Proc. IEEE International Conference on Embedded and Ubiquitous Computing (EUC),* 2017.

[10] M. Modarressi, et al., "Application-Aware Topology Reconfiguration for On-Chip Networks", *IEEE Transactions on Very Large-scale Integrated Circuits and Systems,* Vol. 19, No. 11, pp. 2010-2022, Nov. 2011.

[11] A. Shalaby, et al., "Sentry-NoC: A Statically-Scheduled NoC for Secure SoCs," *in Proc. International Symposium on Networks-on-Chip (NOCS)*, 2021.

[12] G. Wang, et al., "Statistical Analysis and Design of HARP FPGAs", in IEEE Transactions on CAD of Integrated Circuits and Systems, Vol. 25, No. 10, pp. 2088-2102, 2006.

[13] A. Gomez, et al., "Exploiting Wiring Resources on Interconnection Network: Increasing Path Diversity*", in Proc of 16th Euromicro PDP*, 2008.

[14] S. Sahhaf et al., "A Novel SDM-based On-chip Communication Mechanism", in Proc. of European Conference on the Use of Modern Information and Communication Technologies, 2010.

[15] P. Leroy, et al., "Spatial Division Multiplexing: A Novel Approach for Guaranteed Throughput on NoCs", in *Proc. of CODES+ISSS*, pp. 81-86, 2005.

[16] T. Picornell, et al., "DCFNoC: A Delayed Conflict-Free Time Division Multiplexing Network on Chip," in *2019 56th ACM/IEEE Design Automation Conference (DAC)*, 2019.

[17] F. Pakdaman, A. Mazloumi, and M. Modarressi, "Integrated Circuit-Packet Switching NoC with Efficient Circuit Setup Mechanism," *Journal of Supercomputing*, vol. 71, no. 8, pp. 3055–3072, Aug. 2015.

[18] A. Mazloumi, and M. Modarressi, "A hybrid packet/circuit-switched router to accelerate memory access in NoC-based chip multiprocessors," *Design, Automation & Test in Europe Conference & Exhibition (DATE)*, 2015.

[19] M. Gorgues Alonso, J. Flich, M. Turki and D. Bertozzi, "A Low-Latency and Flexible TDM NoC for Strong Isolation in Security-Critical Systems," *2019 IEEE 13th International Symposium on Embedded Multicore/Many-core Systems-on-Chip (MCSoC)*, 2019.

[20] T. Krishna, et al., "SMART: Single-Cycle Multi-Hop Traversals Over A Shared Network-on-Chip", in Special Issue of IEEE Micro, Top Picks from the Computer Architecture Conferences, May/June 2014.

[21] R. Fourer, Robert, D. M. Gay, and B. W. Kernighan, AMPL: A Modeling Language for Mathematical Programming. South SanFrancisco, California: The Scientific Press, 1993.

[22] S. Wang, "Minimizing Power Consumption of Spatial Division Based Networks-on-Chip Using Multipath and Frequency Reduction", *in Proc. of Euromicro DSD*, 2012.

[23] BooksimNoC simulator, http://nocs.stanford.edu/booksim.html.

[24] J. Hu, and R. Marculescu, "Energy- and performance-aware mapping for regular NoC architectures", IEEE Transactions on Computer-Aided Design of Integrated Circuits and Systems, Vol. 24, No. 1, 2005, pp. 551-562.

[25] M. Schmitz, Energy Minimization Techniques for Distributed Embedded Systems, Ph.D. thesis, University of Southampton, 2003.

[26] STG: Standard Task-graph Set, http://www.kasahara.elec.waseda.ac.jp/schedule, June 2014.

[27] Embedded System Synthesis Benchmarks Suite (E3S), http://ziyang.eecs.umich.edu/~dickrp/e3s/, June 2014.